\documentclass[prl,twocolumn,preprintnumbers,amsmath,amssymb,showpacs,nofootinbib,floatfix]{revtex4}

\usepackage{graphicx,bm}

\makeatletter
\def\graphicscale{\twocolumn@sw{0.3}{0.4}}
\def\graphicthreescale{\twocolumn@sw{0.3}{0.4}}

\begin{document}

\title{Finite-size scaling at first-order quantum transitions}

\author{Massimo Campostrini,$^1$ Jacopo Nespolo,$^1$ 
Andrea Pelissetto,$^2$ and Ettore Vicari$^1$} 

\address{$^1$ Dipartimento di Fisica dell'Universit\`a di Pisa
        and INFN, Largo Pontecorvo 3, I-56127 Pisa, Italy}
\address{$^2$ Dipartimento di Fisica di ``Sapienza,'' Universit\`a di Roma
        and INFN, Sezione di Roma I, I-00185 Roma, Italy}

\date{\today}

\begin{abstract}
We study finite-size effects at first-order quantum transitions
(FOQTs). We show that the low-energy properties show a finite-size
scaling (FSS) behavior, the relevant scaling variable being the ratio
of the energy associated with the perturbation driving the transition
and the finite-size energy gap at the FOQT point.  The size dependence
of the scaling variable is therefore essentially determined by the
size dependence of the gap at the transition, which in turn depends on
the boundary conditions.  Our results have broad validity and, in
particular, apply to any FOQT characterized by the degeneracy and
crossing of the two lowest-energy states in the infinite-volume
limit. In this case, a phenomenological two-level theory provides
exact expressions for the scaling functions.  Numerical results for
the quantum Ising chain in transverse and parallel magnetic fields
support the FSS ansatzes.
\end{abstract}


\maketitle



Zero-temperature quantum phase transitions (QPT) are phenomena of
great interest in different branches of
physics~\cite{Sachdev-book,Vojta-03}. They arise in many-body systems
with competing ground states controlled by nonthermal parameters.  QPT
are continuous when the ground state of the system changes
continuously at the transition point and correlation functions show a
divergent length scale. They are instead of first order when the
ground-state properties are discontinuous across the transition point.

These singular behaviors are observed in the infinite-volume limit.
If the size $L$ of the system is finite, all properties are generally
analytic as a function of the external parameter $\mu$ driving the
transition. However, around the transition point, low-energy
thermodynamic quantities and large-scale structural properties show a
finite-size scaling (FSS) behavior depending on the nature and on the
general properties of the transition.  An understanding of these
finite-size properties is important for a correct interpretation of
experimental or numerical data when phase transitions are investigated
in relatively small systems, see, e.g.,
Refs.~\cite{Privman-90,PV-02,GKMD-08}, or in particle systems trapped
by external forces, as in cold-atom experiments, see, e.g.,
Ref.~\cite{BDZ-08}.

At continuous transitions a FSS behavior is observed when the length
scale $\xi$ of the critical modes becomes comparable with the size $L$
of the system.  For large values of $L$, this regime presents
universal features, shared by all systems whose transition belongs to
the same universality class.  Although originally formulated in the
classical framework~\cite{FB-72}, FSS extends to quantum continuous
transitions exploiting the quantum-to-classical
mapping~\cite{SGCS-97,CPV-14}.

First-order quantum transitions (FOQTs) are also of great interest, as
they occur in a large number of quantum many-body systems, such as
quantum Hall samples~\cite{PPBWJ-99}, itinerant
ferromagnets~\cite{VBKN-99}, heavy-fermion
metals~\cite{UPH-04,Pfleiderer-05,KRLF-09}, etc.  FOQTs occur for
those values $\mu_c$ of the external driving parameter $\mu$ at which
the lowest energy states show a level crossing. Since the ground state
is different for $\mu < \mu_c$ and $\mu>\mu_c$, the physical
properties change discontinuously at $\mu_c$.  In the absence of
particular conservation laws, a level crossing can only occur in the
infinite-volume limit. In a finite system, the presence of 
nonvanishing matrix elements among these states lifts the degeneracy,
giving rise to the phenomenon of avoided level crossing.  In this
paper we investigate this issue, in particular we study the interplay
between the size $L$ of the system and the parameter $\mu$ driving the
transition.

First-order {\em classical} transitions (FOCTs), driven by thermal
fluctuations, show FSS behaviors
\cite{NN-75,FB-82,PF-83,FP-85,Privman-90,CLB-86} that are somewhat
analogous to those observed at continuous transitions.  On the basis
of the general quantum-to-classical mapping of $d$-dimensional quantum
systems onto classical anisotropic $(d+1)$-dimensional systems, we
expect FSS also at FOQTs. Here we will show that this is indeed the
case, identifying the general scaling variables which parametrize FSS.

To make the discussion concrete, we start by considering a generic
finite $d$-dimensional cubic system of size $L$ which undergoes a FOQT
driven by a {\em magnetic field} $h$.  We wish to determine the
scaling behavior of the energy differences of the lowest states
$\Delta_n(L,h) = E_n(L,h) - E_0(L,h)$ and of the magnetization
$m(L,h)$.  On dimensional grounds, we expect that around $h=0$ the
relevant {\em scaling} variable is the ratio between the energy
contribution of the magnetic field and the {\em gap} $\Delta_L\equiv
\Delta_1(L,h=0)$ at $h=0$, i.e.,
\begin{equation}
{\kappa}\sim  {hL^d/\Delta_L}.
\label{hvr}
\end{equation}
Then, if a scaling behavior is realized, the energy difference of the
two lowest energy states for finite $h$ should satisfy the scaling
ansatz
\begin{eqnarray}
\Delta(L,h) \equiv \Delta_1(L,h) \approx \Delta_L f_\Delta(\kappa).
\label{deltah}
\end{eqnarray}
We have $f_\Delta(0)=1$ by definition, and $f_\Delta(\kappa)\sim
|\kappa|$ for $\kappa\to\pm \infty$ in order to reproduce the expected
linear behavior $\Delta(L,h) \sim |h|L^d$ for sufficiently large
$|h|$.  The magnetization is expected to scale as
\begin{eqnarray}
m(L,h) \approx m_0 f_\sigma(\kappa),
\label{m0h}
\end{eqnarray}
where $m_0$ is the magnetization obtained approaching the transition
point $h\to 0$ after the infinite-volume limit.  Because of the
definition of $m_0$, we have $f_\sigma(\kappa)\to \pm 1$ for
$\kappa\to \pm \infty$.  Moreover, $f_\sigma(0)=0$ due to the parity
symmetry when $h=0$.  The above FSS ansatzes represent the simplest
scaling behaviors compatible with the discontinuities arising in the
infinite-volume limit. We can also extend the scaling ansatz to
include the temperature $T$.  In the scaling limit $T\to 0$, similar
dimensional arguments indicate that the relevant scaling variable is
the ratio $\rho = T/\Delta_L$, so that, at finite small $T$ we
predict, for instance,
\begin{eqnarray}
m(L,h,T) \approx m_0 F_\sigma(\kappa,\rho).
\label{Dm-scal}
\end{eqnarray}
The scaling variables $\kappa$ and $\rho$ are related to the
finite-size energy gap at the transition, whose size behavior depends
crucially on the boundary conditions considered. Hence, once they are
expressed in terms of the parameter driving the transition and the
size $L$, their $L$ dependence changes according to the chosen
boundary conditions.  As we shall show, it may be proportional to
powers of $L$ as in continuous phase transitions, or be proportional
to exponentials of $L$, as it occurs in Ising classical systems with
cylindrical geometry \cite{PF-83}.

Let us now {\em additionally} assume that the transition is due to the
crossing of the {\em two} lowest-energy states in the infinite-volume
limit.  The higher excited states are instead assumed to be gapped in
the same limit. For example, this occurs in the quantum ferromagnetic
phase of Ising-like systems with appropriate boundary conditions, see
below.  In this case, Eqs.~(\ref{deltah}) and (\ref{m0h}) can be
obtained by a phenomenological theory, which additionally provides the
expressions of the scaling functions.  Because of the hypotheses made,
for sufficiently small $|h|$ the difference between the energies of
the lowest states is much smaller than those between the higher
excited states and the ground state, i.e.
\begin{equation}
\Delta(L,h) \equiv \Delta_1(L,h) \ll \Delta_n(L,h) \quad {\rm
  for}\;\;n>1.
\label{decon}
\end{equation}
Thus, it is natural to assume that the low-energy properties in the
crossover region around $h=0$ are simply obtained by restricting the
theory to the two lowest-energy states.  The Hamiltonian restricted to
this subspace has the general form
\begin{eqnarray}
H_r = \left(
\begin{array}{l@{\ \ }l@{\ \ }l}
\varepsilon  + \beta  &  \quad \delta e^{i\varphi} \\
\delta e^{-i\varphi} &  \quad\varepsilon - \beta\\
\end{array} \right) \; , \label{hr}
\end{eqnarray}
where the basis is chosen so that $\beta \,\sigma^{(3)}$ represents
the perturbation induced by the magnetic field $h$ ($\sigma^{(3)}$ is
the diagonal Pauli matrix), thus $\beta \approx m_0 h L^d$.  $\delta$
is a small parameter which vanishes for $L\to \infty$ and $h=0$, in
order to obtain a degenerate ground state.  Diagonalizing $H_r$, we
obtain the energy difference of the two eigenstates:
\begin{equation}
E_1 - E_0 = 2 \delta \sqrt{1 + (\beta/\delta)^2}.
\label{deem}
\end{equation}
This expression is consistent with the scaling ansatz (\ref{deltah}),
provided that we identify
\begin{equation}
 2\delta \to \Delta_L,\qquad \kappa \equiv {\beta\over \delta} \to  
{2m_0 h L^d\over \Delta_L}, 
\label{corrtbt}
\end{equation}
and
\begin{equation}
f_\Delta(\kappa) = \sqrt{1 + \kappa^2}.
\label{fdelta}
\end{equation}
Moreover, by computing the expectation value of $\sigma^{(3)}$ on the
lowest eigenstate, we obtain that the magnetization satisfies the
scaling relation (\ref{m0h}), with
\begin{equation}
f_\sigma(\kappa) =
{\kappa^2 + \kappa \sqrt{1+\kappa^2} \over 1 + \kappa^2 + 
\kappa \sqrt{1 + \kappa^2}}.
\label{fsigma}
\end{equation}
The FSS functions (\ref{fdelta}) and (\ref{fsigma}) parametrize
$\Delta(L,h)$ and $m(L,h)$ in the limit $L\to\infty$ keeping $\kappa$
fixed. In this limit, scaling corrections are expected to be
suppressed by powers of the inverse size.  Note that the same
technique also provides a framework to study the unitary quantum
dynamics when $h$ varies (in time) in a small interval around $h = 0$.
Since only the two lowest-energy states are relevant, the dynamics is
analogous to that governing a two-level quantum mechanical system in
which the energy separation of the two levels is a function of time
(the Landau-Zener effect~\cite{LZeff}).

In principle, the scaling ansatzes can be derived by exploiting the
general quantum-to-classical mapping, which allows us to map the
quantum system onto a classical one defined in an anisotropic box of
volume $V = L^dL_t$, with $L_t\sim 1/T$ and $L_t\gg L$. We can
therefore check our general predictions using the results of
Ref.~\cite{PF-83} for Ising-like systems with PBC along the 
{\em  transverse} dimensions. In the classical case, FSS is characterized
by the scaling variables $ u \sim hV$ and $ v \sim \xi_t/L_t$, where
$\xi_t$ is the characteristic length at the transition: $\xi_t\sim
\exp (L^d \sigma)$, where $\sigma$ is the interfacial tension.  These
results are fully consistent with our ansatz. If we identify
$\Delta_L$ with $1/\xi_t$, we obtain the correspondence $u/v\to\kappa$
and $v\to\rho$ between the variables $\kappa\sim hL^d/\Delta_L$ and
$\rho \sim T/\Delta_L$ and the classical ones $u$ and $v$.  Note also
that the transfer-matrix analysis of Ref.~\cite{PF-83} is analogous to
the one presented here, as they also observe an avoided-crossing
phenomenon.

\begin{figure}[tbp]
\includegraphics*[scale=\graphicscale]{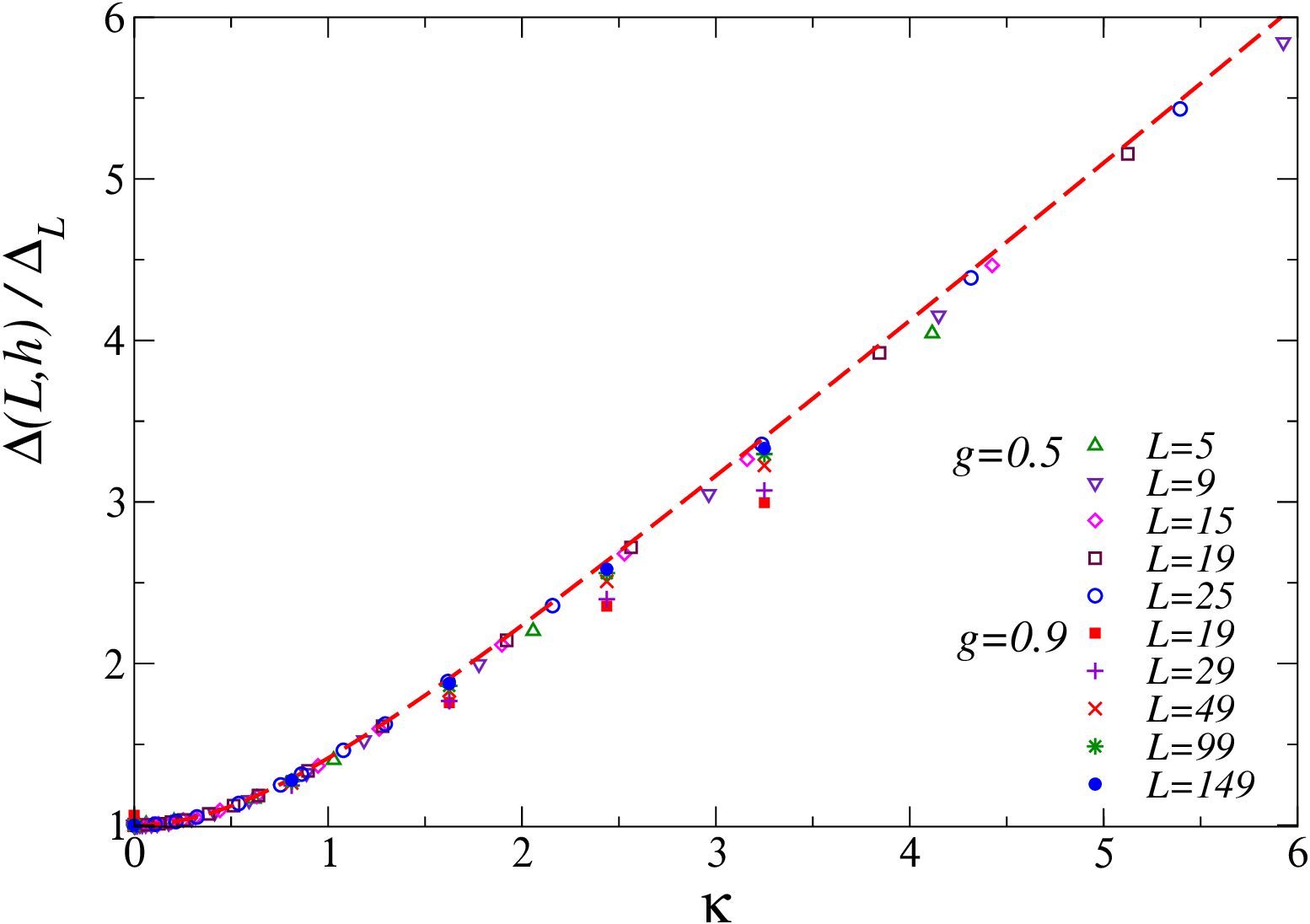}
\includegraphics*[scale=\graphicscale]{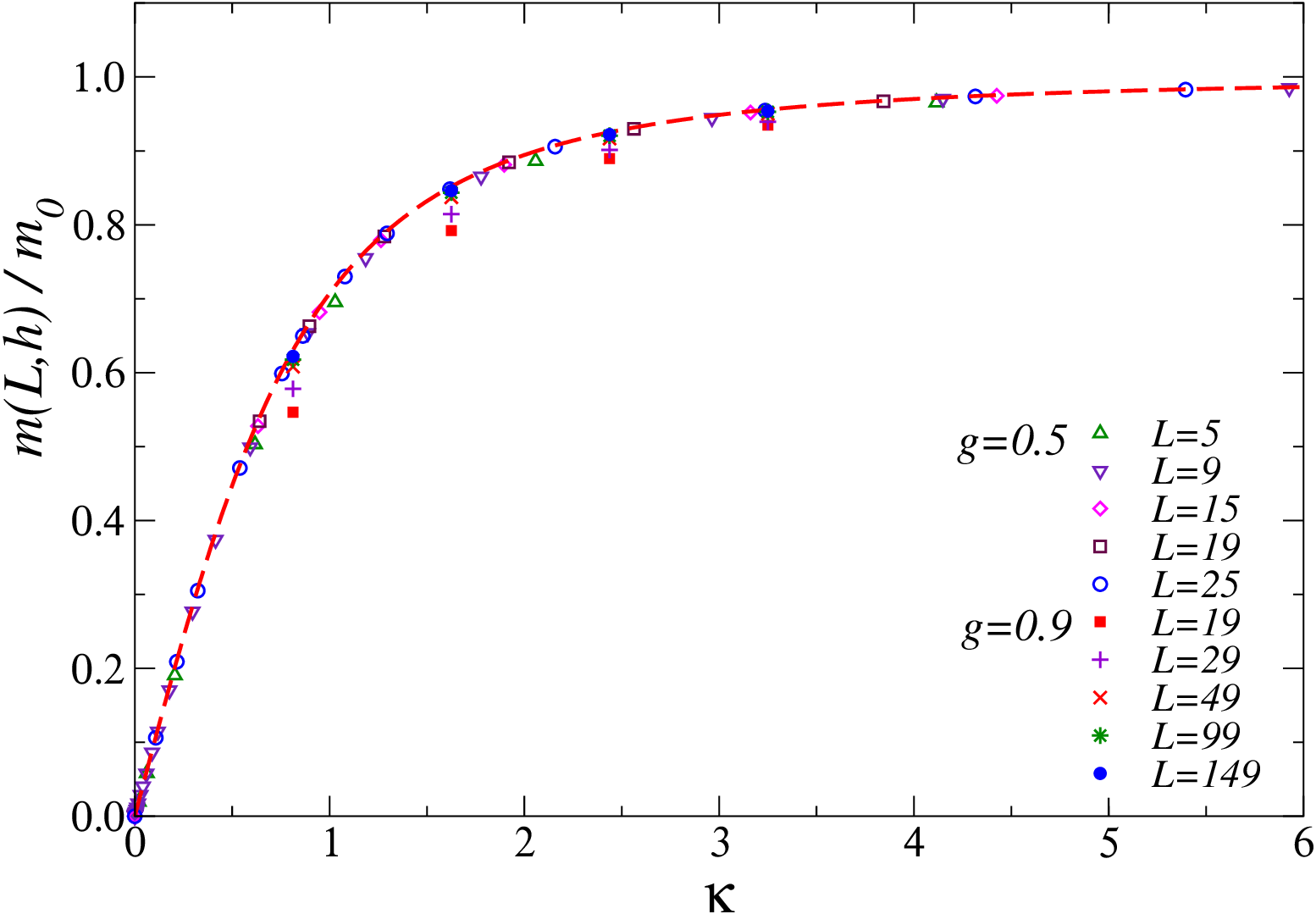}
\caption{(Color online) FSS of the energy difference $\Delta(L,h)$ of
  the lowest states and of the central-site magnetization $m(L,h)$ for
  the Ising chain with OBC, at $g=0.5$ and $g=0.9$.  We plot the
  ratios $\Delta(L,h)/\Delta_L$ (top) and $m(L,h)/m_0$ (bottom) versus
  $\kappa=2m_0 hL/\Delta_L$. We use $m_0=(1-g^2)^{1/8}$ and
  approximate $\Delta_L$ with its leading term, i.e., we set $\Delta_L
  = 2(1-g^2) g^L$ (corrections are small, of order $g^{2L}$).  The
  dashed lines correspond to the scaling functions (\ref{fdelta}) and
  (\ref{fsigma}).  }
\label{desx}
\end{figure}

To verify the general predictions, we consider a quantum Ising chain
of size $L$ in a transverse and parallel magnetic field. Its Hamiltonian 
is
\begin{eqnarray}
H_{\rm Is} = - J \sum_{i} \sigma^{(1)}_i \sigma^{(1)}_{i+1} 
- g \sum_i \sigma^{(3)}_i - h \sum_i \sigma_i^{(1)}, 
\label{Iscd}
\end{eqnarray}
where $\sigma_i^{(a)}$ are the Pauli matrices.  For $h=0$ model
(\ref{Iscd}) has a continuous transition at $g=1$ (we set $J=1$ and
assume $g>0$), belonging to the 2$D$ Ising universality class.  This
quantum critical point separates a paramagnetic ($g>1$) and a
ferromagnetic ($g<1$) phase.  In the ferromagnetic phase, the quantum
Ising chain shows a FOQT at $h=0$ for any $g < 1$, with a
discontinuity of the magnetization, i.e. the ground-state expectation
value of $\sigma_i^{(1)}$.  The Ising chain is an ideal theoretical
laboratory, because several exact results are known for $h=0$.  In the
case of open or periodic boundary conditions (OBC and PBC,
respectively), the transition is due to a crossing of the two
lowest-energy states $| + \rangle$ and $| - \rangle$, such that
\begin{equation}
\langle \pm | \sigma_i^{(1)} | \pm \rangle = \pm \,m_0 
\label{states}
\end{equation}
(this is strictly true for PBC, in the case of OBC boundary effects
are neglected), where~\cite{Pfeuty-70}
\begin{equation}
m_0 = {\rm lim}_{h\to 0^+} {\rm lim}_{L\to\infty} 
\langle \sigma_x^{(1)} \rangle = (1-g^2)^{1/8}.  
\label{modef}
\end{equation} 
However, in a finite system of size $L$, due to tunneling effects, the
lowest states are superpositions of the states $| + \rangle$ and $| -
\rangle$.  Their energy difference $\Delta_L$ vanishes exponentially
as $L$ increases:~\cite{Pfeuty-70,CJ-87}
\begin{eqnarray}
&&\Delta_L \equiv \Delta_1(L,h=0)=2 (1-g^2) g^L\left[1+ O(g^{2L})\right] 
\quad\label{deltalas}
\end{eqnarray}
in the OBC case, and $\Delta_L\approx 2 [(1-g^2)/(\pi L)]^{1/2} g^L$
in the PBC case.  On the other hand, the difference $\Delta_i$ for the
higher excited states ($i > 1$) is finite for $L\to
\infty$~\cite{footnotedelta2}.

Using the density-matrix renormalization-group (DMRG) method
\cite{DMRG,footnoteDMRG}, we compute the energy difference
$\Delta_1(L,h)$ of the lowest two states and the ground-state
magnetization of Ising chains of odd size $L$, such that $-L/2\le i
\le L/2$, with OBC.  In the case of OBC there is no translation
invariance, thus $M_i=\langle \sigma_i^{(1)}\rangle$ depends on
$i$. Such a dependence is weak far from the boundary.  In the
following, we consider $m(L,h) \equiv M_0(L,h)$ at the center of the
lattice.  Fig.~\ref{desx} shows data at $g=0.5$ and $g=0.9$.  They
clearly show the predicted FSS behavior. Indeed, the ratios
$\Delta(L,h)/\Delta_L$ and $m(L,h)/m_0$ approach universal, $g$
independent, curves when they are plotted versus $\kappa=2m_0
hL/\Delta_L$.  Moreover, the scaling functions agree with those
given by the phenomenological two-level theory, i.e., with
Eqs.~(\ref{fdelta}) and (\ref{fsigma}).  These behaviors appear to set
in for quite small lattice sizes, with corrections that turn out to
asymptotically decrease as $O(1/L)$.  

As an additional check of the general argument, we consider systems in
which the parallel magnetic field $h_j$ is nonvanishing only at one
lattice site $j$, for example at the center of the chain.  The general
arguments should apply to this case as well, with $\beta\sim h_j$
instead of $\beta\sim hL$.  Therefore, we expect a FSS behavior
analogous to that valid for the homogenous parallel field, the
corresponding scaling variable being $\kappa = 2 m_0 h_j/\Delta_L$.
DMRG results for the Ising chain are in full agreement with these
scaling predictions.  In particular, the scaling behavior of
$\Delta(L,h)$ given by Eqs.~(\ref{deltah}) and (\ref{fdelta}) can
be analytically derived for magnetic fields localized at the
boundaries. These calculations will be reported elsewhere.

Although our discussion has essentially focused on Ising-like
systems, the two-level theory is general and applies to any FOQT in
which two levels cross in the infinite-volume limit. Indeed, the
parity symmetry $h\to -h$ plays no role here, as it can be easily
checked by allowing for generic diagonal terms in the matrix
(\ref{hr}), i.e. by replacing $\epsilon\pm\beta$ with
$\epsilon+\beta_1$ and $\epsilon+\beta_2$, where $\beta_1,\,\beta_2$
represent the energy contributions of the driving perturbation.
Moreover, it can be easily extended to systems with a larger finite
degeneracy at the transition, keeping into account a larger number of
low-energy states.

\begin{figure}[tbp]
\includegraphics*[scale=\graphicscale]{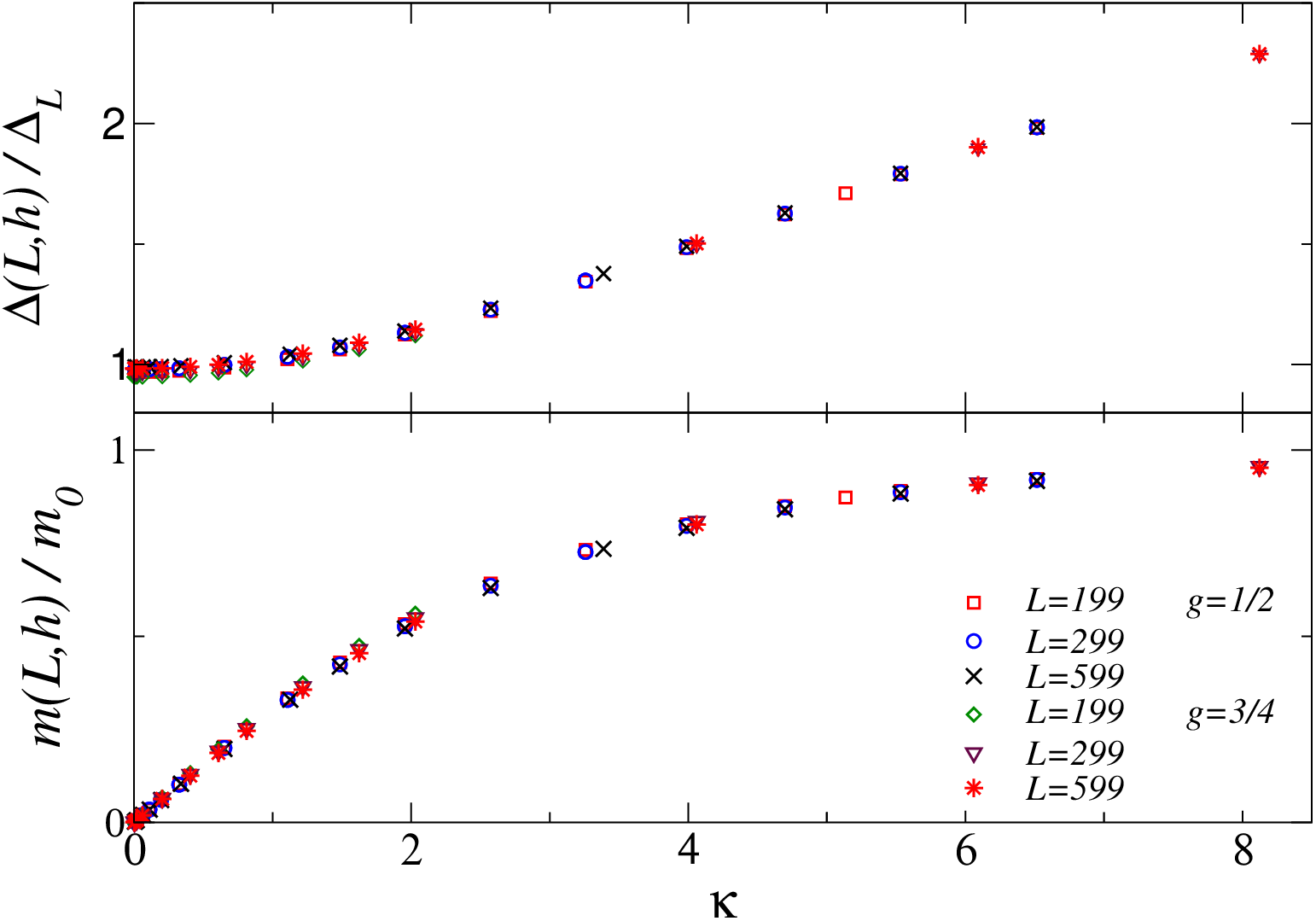}
\caption{(Color online) FSS of the energy difference of the lowest
  states and magnetization of the Ising chain with FOBC, at $g=1/2$
  and $g=3/4$.  We plot the ratios $\Delta(L,h)/\Delta_L$ (top) and
  $m(L,h)/m_0$ (bottom) versus $\kappa \equiv 2 m_0 h L/\Delta_L$,
  with $m_0$ and $\Delta_L$ given by Eqs.~(\ref{modef}) and (\ref{deltaabcfobc}).
  The data approach nontrivial scaling curves
  with increasing $L$, which are independent of $g$,
supporting the FSS Eqs.~(\ref{deltah}) and (\ref{m0h}).  }
\label{deosx}
\end{figure}

Finally, we show that the FSS Eqs.~(\ref{deltah}) and (\ref{m0h}) also
hold in systems with an infinite degeneracy of the ground state at the
transition.  To investigate this case, we consider the Ising chain
with antiperiodic boundary conditions (ABC), or fixed opposite
boundary conditions (FOBC)~\cite{footnotefobc}.  In these two cases,
the lowest-energy states are associated with domain walls (kinks),
i.e., with nearest neighbors pairs of antiparallel spins, which can be
considered as one-particle states with $O(L^{-1})$ momenta. Hence,
there is an infinite number of excitations with a gap of order
$L^{-2}$.  In particular, we have
\begin{equation}
\Delta_L \equiv \Delta_1(L,h=0)
= c \,{g\over 1-g} \, {\pi^2\over L^2} + O(L^{-3})
\label{deltaabcfobc}
\end{equation}
with $c=1$ for ABC~\cite{CJ-87} and $c=3$ for FOBC.  DMRG calculations
show that scaling ansatzes (\ref{deltah}) and (\ref{m0h}) hold also in
this case.  Indeed, like the OBC case, the data for the ratios
$\Delta(L,h)/\Delta_L$ and $m(L,h)/m_0$ approach universal
(independent of $g$) scaling curves when plotted versus $\kappa \equiv
2 m_0 h L/\Delta_L$, see Fig.~\ref{deosx}.

Note that, if we express the scaling variable $\kappa$ in terms of
$L$, the $L$ dependence is quite different from that observed in the
OBC and PBC case. While $\kappa \sim h L g^{-L}$ and $\kappa\sim h
L^{3/2} g^{-L}$ for OBC and PBC, respectively, we have $\kappa \sim h
L^3$ for ABC and FOBC, since $\Delta_L\sim L^{-2}$,
cf. Eq.~(\ref{deltaabcfobc}).  Including the temperature 
in the case of ABC and FOBC, we obtain the scaling variable $\rho = T
L^2$, which shows that the FSS limit in the corresponding classical
system is strongly anisotropic (the same occurs in the PBC case,
already considered in Ref.~\cite{PF-83}).  These results show that the
FSS dependence may significantly depend on the boundary conditions,
being strictly connected with the large-volume low-energy structure of
the eigenstates.

In conclusion, quantum systems show a universal FSS behavior at
FOQTs. For instance, the gap satisfies Eq.~(\ref{deltah}) when the
scaling variable $\kappa$ is expressed in terms of the ratio of the
energy associated with the perturbation driving the transition and the
finite-size energy gap at the FOQT. We show that these results hold
for general FOQTs characterized by the degeneracy and crossing of the
two lowest-energy states in the infinite-volume limit, and that, in
this case, it is possible to predict the scaling functions.  We also
show that the same FSS predictions also hold in some systems in which
the degeneracy of the ground state at the transition is infinite (this
is the case of the Ising chain with ABC or FOBC), indicating that
Eq.~(\ref{deltah}) has a broad range of validity.  It is important to
note that $\kappa$ depends on $\Delta_L$, whose $L$ dependence, in
turn, varies with the boundary conditions. This implies that $\kappa$
may show different $L$ dependences for different boundary conditions.
For instance, for the Ising chain $\kappa \sim h L e^{\alpha L}$ for
OBC, $\kappa \sim h L^{3/2} e^{\alpha L}$ for PBC ($\alpha$ is a
nonuniversal positive constant), and $\kappa \sim h L^3$ for ABC or
FOBC.  Further work is needed to investigate systems with vector
symmetries, which, in the classical case \cite{FP-85}, have a more
complex behavior due to the presence of massless Goldstone bosons.

The FSS theory at FOQTs should be particularly useful in those cases
in which the nature of the transition is not known {\em a priori} and
only data for relatively small systems are available: FSS may be
exploited to discriminate continuous from weak FOQTs.  We also note
that, in some cases, FSS behaviors are already observed for relatively
small systems, for example in the case of Ising chains with OBC, see
Fig.~\ref{desx}.  Thus, even small systems may show definite
signatures of FOQTs, as also argued in Refs.~\cite{IZ-04,LMD-11}. This
makes FSS at FOQTs particularly interesting also experimentally.  In
particular, quantum simulators \cite{GAN-14}, in which an extremely
high degree of control is achieved in simple systems of a few atoms,
have already been used to investigate the quantum critical behavior of
different models \cite{LMD-11, Islam-etal-11, Simon-etal-11}, and
could provide experimental support to the scaling theory introduced in
the present work.


\begin{thebibliography}{99}
\bibitem{footnoteDMRG} Since the system is always far from
  criticality, DMRG is extremely efficient; however, for very small
  energy differences, computation of eigenvectors and therefore of
  vacuum expectation values can be inaccurate (this is true of any
  numerical diagonalization method).  The precision on the observables
  is typically $ 10^{-15} |E_0|/\Delta_L$ with $|E_0| \simeq L$.
\end{thebibliography}
\end{document}